\documentclass[12pt,english]{article}
\usepackage{babel}
\usepackage{amsfonts,epsfig}
\title{\bf Bogoliubov compensation principle in the electro-weak interaction:
value of the gauge constant, muon g-2 anomaly, predictions for Tevatron and LHC}
\author{{\bf Boris A. Arbuzov}\\
{Skobeltsyn Institute of Nuclear Physics of
MSU,}\\ {119992 Moscow, Russia}
\\
\\
Bogoliubov compensation principle in the EW interaction }

\date{}
\begin{document}
\maketitle
\begin{quote}
We apply Bogoliubov compensation principle to the gauge electro-weak interaction
to demonstrate a  spontaneous generation of anomalous three-boson gauge invariant
effective interaction. The non-trivial solution of compensation equations uniquely defines values of parameters of the theory and the form-factor of the anomalous interaction. The contribution of this interaction to
running EW coupling $\alpha_{ew}(p^2)$ gives its observable value
$\alpha_{ew}(M_W^2)=0.0374$ in satisfactory agreement to the experiment. The anomalous three-boson interaction gives natural explanation of the well-known discrepancy in muon
$g-2$. The implications for EW studies at Tevatron and LHC are briefly discussed.

\end{quote}

\section{Introduction}

In previous works~\cite{Arb04, Arb05, AVZ, Arb07, AVZ2}
N.N. Bogoliubov compensation principle~\cite{Bog1, Bog2, Bog} was applied to studies of spontaneous generation of effective non-local interactions in renormalizable
gauge theories.  Spontaneous generation of Nambu -- Jona-Lasinio like
interaction was studied in works~\cite{Arb05, AVZ, AVZ2} and the description of low-energy
hadron physics in terms of initial QCD parameters turns to be quite successful including values
of parameters: $m_\pi,\, f_\pi,\, m_\sigma,\,
\Gamma_\sigma,\, <\bar q q>, M_\rho, \Gamma_\rho, M_{a_1}, \Gamma_{a_1}$.

In work~\cite{Arb07}  the approach was applied for calculation of infrared
behaviour of the QCD running coupling constant. In particular, in QCD a possibility of spontaneous generation of
anomalous three-gluon interaction of the form
\begin{equation}
-\,\frac{G}{3!}\cdot\,f_{abc}\,F_{\mu\nu}^a\,F_{\nu\rho}^b\,F_{\rho\mu}^c\,;
\label{FFF}
\end{equation}
was shown.

The main principle of the approach is to check if an effective interaction
could be generated in a chosen variant of a renormalizable theory. In view
of this one performs "add and subtract" procedure for the effective
interaction with a form-factor. Then one assumes the presence of the
effective interaction in the interaction Lagrangian and the same term with
the opposite sign is assigned to the newly defined free Lagrangian. This
transformation of the initial Lagrangian is evidently identical. However
such free Lagrangian contains completely improper term, corresponding to
the effective interaction of the opposite sign. Then one
has to formulate a compensation equation, which guarantees that this new
free Lagrangian is a genuine free one, that is effects of the
uncommon term sum up to zero. Provided a non-trivial solution of this
equation exists, one can state the generation of the effective
interaction to be possible. Now we apply this procedure to our problem.

In the present work we consider a possibility of generation of
interaction analogous to~(\ref{FFF}) in the electro-weak theory.

\section{Compensation equation for anomalous three-boson interaction}

We start with EW Lagrangian with $3$ lepton and colour quark doublets
with gauge group $SU(2)$. That is we restrict the gauge sector to
triplet of $W^a_\mu$ only. Thus we consider $U(1)$ abelian gauge field $B$ to be decoupled, that means approximation $\sin^2\theta_W \ll 1$.
\begin{eqnarray}
& & L\,=\,\sum_{k=1}^3\biggl(\frac{\imath}{2}
\Bigl(\bar\psi_k\gamma_
\mu\partial_\mu\psi_k\,-\partial_\mu\bar\psi_k
\gamma_\mu\psi
_k\,\biggr)-\,m_k\bar\psi_k\psi_k\,+
\,\frac{g}{2}\,\bar\psi_k\gamma_\mu \tau^a W_\mu^a\psi_k
\biggr)\,+\label{initial}\\
& & +\sum_{k=1}^3\biggl(\frac{\imath}{2}
\Bigl(\bar q_k\gamma_
\mu\partial_\mu q_k\,-\partial_\mu\bar q_k
\gamma_\mu q_k\,\biggr)-\,M_k\bar q_k q_k\,+
\,\frac{g}{2}\,\bar q_k\gamma_\mu \tau^a W_\mu^a q_k
\biggr)\,-\nonumber\\
& &-\frac{1}{4}\,\biggl( W_{\mu\nu}^aW_{\mu\nu}^a\biggr);\qquad
W_{\mu\nu}^a\,=\,
\partial_\mu W_\nu^a - \partial_\nu W_\mu^a\,+g\,\epsilon_{abc}W_\mu^b W_\nu^c\,.
\nonumber
\end{eqnarray}
where we use the standard notations and $\psi_k$ and $q_k$ correspond to leptons and quarks respectfully.
In accordance to the Bogoliubov approach~\cite{Bog1, Bog2, Bog} in application to
QFT~\cite{Arb04} we look for
a non-trivial solution of a
compensation equation, which is formulated on the basis
of the Bogoliubov procedure {\bf add -- subtract}. Namely
let us write down the initial expression~(\ref{initial})
in the following form
\begin{eqnarray}
& &L\,=\,L_0\,+\,L_{int}\,;\nonumber\\
& &L_0\,=\,=\,\sum_{k=1}^3\biggl(\frac{\imath}{2}
\Bigl(\bar\psi_k\gamma_
\mu\partial_\mu\psi_k\,-\partial_\mu\bar\psi_k
\gamma_\mu\psi
_k\,\biggr)-\,m_k\bar\psi_k\psi_k\,+\frac{\imath}{2}
\Bigl(\bar q_k\gamma_
\mu\partial_\mu q_k\,-\partial_\mu\bar q_k
\gamma_\mu q_k\,\biggr)-\nonumber\\
& &-\,M_k\bar q_k q_k\biggr)\,-\,\frac{1}{4}\,W_{\mu\nu}^a W_{\mu\nu}^a\,+
\,\frac{G}{3!}\cdot\,\epsilon_{abc}\,W_{\mu\nu}^a\,W_{\nu\rho}^b\,W_{\rho\mu}^c\,;
\label{L0}\\
& &L_{int}\,=\,\frac{g}{2}\,\sum_{k=1}^3\biggl(\bar\psi_k\gamma_\mu
\tau^a W_\mu^a\psi_k\,+\,\bar q_k\gamma_\mu \tau^a W_\mu^a q_k\biggr)\,-\,\frac{G}{3!}\cdot\,\epsilon_{abc}\,
W_{\mu\nu}^a\,W_{\nu\rho}^b\,W_{\rho\mu}^c\,.\label{Lint}
\end{eqnarray}

Here isotopic summation is
performed inside of each quark
bi-linear combination, and notation
 $-\,\frac{G}{3!}\cdot \,\epsilon_{abc}\,
W_{\mu\nu}^a\,W_{\nu\rho}^b\,W_{\rho\mu}^c$ means corresponding
non-local vertex in the momentum space
\begin{eqnarray}
& &(2\pi)^4\,G\,\,\epsilon_{abc}\,(g_{\mu\nu} (q_\rho pk - p_\rho qk)+ g_{\nu\rho}
(k_\mu pq - q_\mu pk)+g_{\rho\mu} (p_\nu qk - k_\nu pq)+\nonumber\\
& &+\,q_\mu k_\nu p_\rho - k_\mu p_\nu q_\rho)\,F(p,q,k)\,
\delta(p+q+k)\,+...;\label{vertex}
\end{eqnarray}
where $F(p,q,k)$ is a form-factor and
$p,\mu, a;\;q,\nu, b;\;k,\rho, c$ are respectfully incoming momenta,
Lorentz indices and weak isotopic indices
of $W$-bosons. We mean also that there are present four-boson, five-boson and
six-boson vertices according to expression for $W_{\mu\nu}^a$
(\ref{initial}).

Effective interaction
\begin{equation}
-\,\frac{G}{3!}\cdot \,\epsilon_{abc}\,
W_{\mu\nu}^a\,W_{\nu\rho}^b\,W_{\rho\mu}^c\,\label{effint}
\end{equation}
 is
usually called {\bf anomalous three-boson interaction} and it is
considered for long time on phenomenological grounds~\cite{Hag}.
Note, that the first attempt to obtain the anomalous three-boson
interaction in the framework of Bogoliubov approach was done in work
~\cite{Arb92}. Our interaction constant $G$ is connected with
conventional definitions in the following way
\begin{equation}
G\,=\,\frac{g\,\lambda}{M_W^2}\,.\label{Glam}
\end{equation}
The current limitations for parameter $\lambda$ read~\cite{EW}
\begin{eqnarray}
& &\lambda\,=\,-\,0.016^{+0.021}_{-0.023}\,;\qquad -\,0.059< \lambda < 0.026\,
(95\%\,C.L.)\,;
\nonumber\\
& &\lambda\,=\,-\,0.024^{+0.025}_{-0.021}\,;\qquad -\,0.068< \lambda < 0.023\,
(95\%\,C.L.)\,;\label{lambda1}
\end{eqnarray}
where the first line corresponds to one-parameter fit ($\lambda$) with zero anomalous electric quadrupole moment of $W$-boson $\kappa$, while the second line corresponds  to two-parameter fit, including also $\kappa$. Due to our approximation $\sin^2\theta_W\,
\ll\,1$ we use the same $M_W$ for both charged $W^\pm$ and neutral $W^0$ bosons
and assume no difference in anomalous interaction for $Z$ and $\gamma$, i.e.
$\lambda_Z\,=\,\lambda_\gamma\,=\,\lambda$.

Let us consider  expression~
(\ref{L0}) as the new {\bf free} Lagrangian $L_0$,
whereas expression~(\ref{Lint}) as the new
{\bf interaction} Lagrangian $L_{int}$. It is important to note, that we
put into the new {\bf free} Lagrangian the full quadratic in $W$ term including
self-boson interaction, because we prefer to maintain gauge invariance of the approximation being used. Indeed, we shall use both quartic term from the last term
in~(\ref{L0}) and triple one from the last but one term of~(\ref{L0}).
Then compensation conditions (see for details~\cite{Arb04}) will
consist in demand of full connected three-boson vertices of the structure~(\ref{vertex}),
following from Lagrangian $L_0$, to be zero. This demand
gives a non-linear equation for form-factor $F$.

Such equations according to terminology of works
~\cite{Bog1, Bog2, Bog} are called {\bf compensation equations}.
In a study of these equations it is always evident the
existence of a perturbative trivial solution (in our case
$G = 0$), but, in general, a non-perturbative
non-trivial solution may also exist. Just the quest of
a non-trivial solution inspires the main interest in such
problems. One can not succeed in finding an exact
non-trivial solution in a realistic theory, therefore
the goal of a study is a quest of an adequate
approach, the first non-perturbative approximation of
which describes the main features of the problem.
Improvement of a precision of results is to be achieved
by corrections to the initial first approximation.

Thus our task is to formulate the first approximation.
Here the experience acquired in the course of performing
works~\cite{Arb04, Arb05, AVZ, Arb07} could be helpful. Now in view of
obtaining the first approximation we would make the following
assumptions.\\
1) In compensation equation we restrict ourselves by
terms with loop numbers 0, 1.\\
2) We reduce thus obtained non-linear compensation equation to a linear
integral equation. It means that in loop terms only one vertex
contains the form-factor, being defined above, while
other vertices are considered to be point-like. In
diagram form equation for form-factor $F$ is presented
in Fig. 1. Here four-leg vertex correspond to interaction of four
bosons due to our effective three-field interaction. In our approximation we
take here point-like vertex with interaction constant proportional
to $g\,G$.\\
3) We integrate by angular variables of the 4-dimensional Euclidean
space. The necessary rules are presented in paper~\cite{Arb05}.

At first let us present the expression for four-boson vertex
\begin{eqnarray}
& &V(p,m,\lambda;\,q,n,\sigma;\,k,r,\tau;\,l,s,\pi) = g G \biggl(\epsilon^{amn}
\epsilon^{ars}\Bigl(U(k,l;\sigma,\tau,\pi,\lambda)-U(k,l;\lambda,\tau,\pi,\sigma)-\nonumber\\& &-U(l,k;\sigma,\pi,\tau,\lambda)+
U(l,k;\lambda,\pi,\tau,\sigma)+U(p,q;\pi,\lambda,\sigma,\tau)-U(p,q;\tau,\lambda,\sigma,\pi)-\nonumber\\
& &-U(q,p;\pi,\sigma,\lambda,\tau)
+U(q,p;\tau,\sigma,\lambda,\pi)\Bigr)-\epsilon^{arn}\,
\epsilon^{ams}\Bigl(U(p,l;\sigma,\lambda,\pi,\tau)-\nonumber\\
& &-U(l,p;\sigma,\pi,\lambda,\tau)
-U(p,l;\tau,\lambda,\pi,\sigma)+
U(l,p;\tau,\pi,\lambda,\sigma)+U(k,q;\pi,\tau,\sigma,\lambda)-\nonumber\\
& &-U(q,k;\pi,\sigma,\tau,\lambda)
-U(k,q;\lambda,\tau,\sigma,\pi)
+U(q,k;\lambda,\sigma,\tau,\pi)\Bigr)+\label{four}\\
& &+\epsilon^{asn}\,
\epsilon^{amr}\Bigl(U(k,p;\sigma,\lambda,\tau,\pi)-U(p,k;\sigma,\tau,\lambda,\pi)
+U(p,k;\pi,\tau,\lambda,\sigma)-\nonumber\\
& &-U(k,p;\pi,\lambda,\tau,\sigma)-U(l,q;\lambda,\pi,\sigma,\tau)
+U(l,q;\tau,\pi,\sigma,\lambda)
-U(q,l;\tau,\sigma,\pi,\lambda)+\nonumber\\
& &+U(q,l;\lambda,\sigma,\pi,\tau)\Bigr)\biggr)\,;\nonumber\\
& &U(k,l;\sigma,\tau,\pi,\lambda)=k_\sigma\,l_\tau\,g_{\pi\lambda}-k_\sigma\,l_\lambda\,g_{\pi\tau}+k_\pi\,l_\lambda\,g_{\sigma\tau}-(kl)g_{\sigma\tau}g_{\pi\lambda}\,.\nonumber
\end{eqnarray}
Here triad $p,\,m,\,\lambda$ {\it etc} means correspondingly incoming momentum, isotopic
index, Lorentz index of a boson.

Let us formulate compensation equations in this
approximation.
For {\bf free} Lagrangian $L_0$ full connected
three-boson vertices with Lorentz structure~(\ref{vertex}) are to vanish. One can succeed in
obtaining analytic solutions for the following set
of momentum variables (see Fig. 1): left-hand legs
have momenta  $p$ and $-p$, and a right-hand leg
has zero momentum.
However in our approximation we need form-factor $F$ also
for non-zero values of this momentum. We look for a solution
with the following simple dependence on all three variables
\begin{equation}
F(p_1,\,p_2,\,p_3)\,=\,F(\frac{p_1^2\,+\,p_2^2\,+\,p_3^2}{2})\,;\label{123}
\end{equation}
Really, expression~(\ref{123}) is symmetric and it turns to $F(x)$
for $p_3=0,\,p_1^2\,=\,p_2^2\,=\,x$. We consider the representation~(\ref{123})
to be the first approximation and we plan to take into account the
corresponding correction in forthcoming studies.

Now according to the rules being stated above we
obtain the following equation for form-factor $F(x)$
\begin{eqnarray}
& &F(x)\,=\,-\,\frac{G^2\,N}{64\,\pi^2}\Biggl(\int_0^Y\,F(y)\,y dy\,-\,
\frac{1}{12\,x^2}\,\int_0^x\,F(y)\,y^3 dy\,
+\,\frac{1}{6\,x}\,\int_0^x\,F(y)\,
y^2 dy\,+\nonumber\\
& &+\,\frac{x}{6}\,\int_x^Y\,F(y)\,dy\,-\,\frac{x^2}{12}\,
\int_x^Y\,\frac{F(y)}{y}\,dy \Biggr)\,+\,\frac{G\,g\,N}{16\,\pi^2}\,
\int_0^Y\,F(y)\, dy\,+\label{eqF}\\
& &+\,\frac{G\,g\,N}{24\,\pi^2}\,\Biggl(\int_{3 x/4}^{x}\,\frac{(3 x-  4 y)^2 (3 x- 8 y)}{x^2 (x-2 y)}F(y)\,
dy\,+\,\int_{x}^Y\,\frac{(5 x- 6 y)}{(x-2 y)}F(y)\, dy\Biggr)\,+\nonumber\\
& &+\,\frac{G\,g\,N}{32 \pi^2}\Biggl(\int_{3 x/4}^{x}\,\frac{3(4 y-3 x)^2(x^2-4 x y+2 y^2)}{8 x^2(2 y-x)^2}\,F(y)\,dy\,+\,\int_x^Y\,\frac{3(x^2-2 y^2)}{8(2 y-x)^2}\,F(y)\,dy
\,+\nonumber\\
& &+\,\int_0^x\frac{5 y^2-12 x y}{16 x^2}\,F(y)\,dy\,+\,\int_x^Y\,
\frac{3 x^2- 4 x y - 6 y^2}{16 y^2}\,F(y)\,dy\Biggr)\,.\nonumber
\end{eqnarray}
Here $x = p^2$ and $y = q^2$, where $q$ is an integration momentum, $N=2$. The last four terms in brackets represent diagrams with one usual gauge vertex (see three last
diagrams at Fig. 1). We introduce here
an effective cut-off $Y$, which bounds a "low-momentum" region where
our non-perturbative effects act
and consider the equation at interval $[0,\, Y]$ under condition
\begin{equation}
F(Y)\,=\,0\,. \label{Y0}
\end{equation}
We shall solve equation~(\ref{eqF}) by iterations. That is we
expand its terms being proportional to $g$ in powers of $x$ and
take at first only constant term. Thus we have
\begin{eqnarray}
& &F_0(x)\,=\,-\,\frac{G^2\,N}{64\,\pi^2}\Biggl(\int_0^Y\,F_0(y)\,y dy\,-\,
\frac{1}{12\,x^2}\,\int_0^x\,F_0(y)\,y^3 dy\,
+\,\frac{1}{6\,x}\,\int_0^x\,F_0(y)\,
y^2 dy\,+\nonumber\\
& &+\,\frac{x}{6}\,\int_x^Y\,F_0(y)\,dy\,-\,\frac{x^2}{12}\,
\int_x^Y\,\frac{F_0(y)}{y}\,dy \Biggr)\,+\,\frac{87\,G\,g\,N}{512\,\pi^2}\,\int_0^Y\,F_0(y)\, dy\,.\label{eqF0}
\end{eqnarray}
Expression~(\ref{eqF0}) provides an equation of the type which were
studied in papers~\cite{Arb04, Arb05, AVZ, Arb07},
where the way of obtaining
solutions of equations analogous to (\ref{eqF0}) are described.
Indeed, by successive differentiation of Eq.(\ref{eqF0}) we come to
Meijer differential equation~\cite{be}
\begin{eqnarray}
& &\biggl(x\,\frac{d}{dx} + 2\biggr)\biggl(x\,\frac{d}{dx} + 1\biggr)\biggl(x\,\frac{d}{dx} - 1\biggr)\biggl(x\,\frac{d}{dx} - 2\biggr)F(x)\,+
\,\frac{G^2\,N\,x^2}{64\,\pi^2}\,F(x)\,=\label{difur}\\
& &=\,4\,\Biggl(-\,\frac{G^2\,N}{64\,\pi^2}\,\int_0^Y F_0(y)
\,y dy + \frac{87\,G\,g\,N}{512\,\pi^2}\,\int_0^Y F_0(y)\, dy
\Biggr)\,;\nonumber
\end{eqnarray}
which solution looks like
\begin{eqnarray}
& &F_0(z) = \,C_1\,G_{04}^{10}\Bigl( z\,|1/2,\,1,\,-1/2,\,-1\Bigr) +
C_2\,G_{04}^{10}\Bigl( z\,|1,\,1/2,\,-1/2,\,-1\Bigr)\,- \label{solution}\\
& &-\,\frac{G\,N}{128\,\pi^2}\,G_{15}^{31}\Bigl( z\,|^0_{1,\,1/2,\,0,\,-1/2,\,-1}\Bigr)\, \int_0^Y \Biggl(G \,y\,-\,\frac{87\, g}{8}\Biggr)F_0(y)\,dy\,
;\nonumber\\
& & G_{15}^{31}\Bigl( z\,|^0_{1,\,1/2,\,0,\,-1/2,\,-1}\Bigr)=
\frac{1}{2\,z}-G_{04}^{30}\Bigl( z\,|1,\,1/2,\,-1,\,-1/2\Bigr)\,;\quad
z\,=\,\frac{G^2\,N\,x^2}{1024\,\pi^2}\,;\nonumber
\end{eqnarray}
where
$$
G_{qp}^{nm}\Bigl( z\,|^{a_1,..., a_q}_{b_1,..., b_p}\Bigr)\,;
$$
is a Meijer function~\cite{be}. In case $q=0$ we write only indices $b_i$ in one
line. Constants $C_1,\,C_2$ are defined by the following boundary conditions
\begin{eqnarray}
& &\Bigl[2\,z^2 \frac{d^3\,F_0(z)}{dz^3}\,+9\,z\,\frac{d^2\,F_0(z)}{dz^2}\,+\,
\frac{d\,F_0(z)}{dz}\Bigr]_{z\,=\,z_0} = 0\,;\nonumber\\
& &\Bigl[2\,z^2\,\frac{d^2\, F_0(z)}{dz^2}\,+5\,z\,\frac{d\, F_0(z)}{dz}\,+\,
F_0(z) \Bigr]_{z\,=\,z_0} = 0\,;
\quad z_0\,=\,\frac{G^2\,N\,Y^2}{1024\,\pi^2}\,.
\label{bc}
\end{eqnarray}

Conditions~(\ref{Y0}, \ref{bc}) defines set of
parameters
\begin{equation}
z_0\,=\,\infty\,; \quad C_1\,=\,0\,
; \quad C_2\,=\,0\,.\label{z0C}
\end{equation}
The normalization condition for form-factor $F(0)=1$ here is the following
\begin{equation}
-\,\frac{G^2\,N}{64\,\pi^2}\,\int_0^\infty F_0(y)
\,y
dy + \frac{87\,G\,g\,N}{512\,\pi^2} \int_0^\infty F_0(y)\,dy\, =\,1\,.
\label{norm}
\end{equation}
However the first integral in (\ref{norm}) diverges due to asymptotics
$$
G_{15}^{31}\Bigl( z\,|^0_{1,\,1/2,\,0,\,-1/2,\,-1}\Bigr)\,\to\,
\frac{1}{2\,z}\,, \quad z\,\to\,\infty\,;
$$
and we have no consistent solution. In view of this we consider the next
approximation. We substitute solution (\ref{solution}) with account of~(\ref{norm}) into terms of Eq.~(\ref{eqF}) being proportional to gauge constant $g$ but the
constant ones and
calculate terms  proportional to $\sqrt{z}$. Now we have bearing in mind the normalization condition
\begin{eqnarray}
& &F(z)\,=\,1 + \frac{85\, g\,\sqrt{N} \,\sqrt{z}}{96\,\pi}\Biggl(
\ln\,z + 4\,
\gamma + 4\,\ln\,2 +\frac{1}{2}\,G_{15}^{31}\Bigl( z_0\,|^0_{0, 0, 1/2, -1, -1/2}\Bigr) - \nonumber\\
& &-\,\frac{3160}{357}\Biggr) +
\frac{2}{3\,z} \int_0^z F(t)\,t\, dt - \frac{4}{3\,\sqrt{z}} \int_0^z F(t)
\sqrt{t}\, dt - \frac{4\,\sqrt{z}}{3} \int_z^{z_0} F(t) \frac{dt}{\sqrt{t}}\,+
\nonumber\\
& &+\,\frac{2\,z}{3}\,\int_z^{z_0}\,F(t)\,\frac{dt}{t}\,;\label{eqFg}
\end{eqnarray}
where $\gamma$ is the Euler constant. We look for solution of (\ref{eqFg})
in the form
\begin{eqnarray}
& &F(z)\,=\,\frac{1}{2}\,G_{15}^{31}\Bigl( z\,|^0_{1,\,1/2,\,0,\,-1/2,\,-1}
\Bigr) -\,\frac{85\,g \sqrt{N}}{512\,\pi}\,G_{15}^{31}\Bigl( z\,|^{1/2}_{1,\,1/2,
\,1/2,\,-1/2,\,-1}\Bigr)\,+\nonumber\\
& &+\,C_1\,G_{04}^{10}\Bigl( z\,|1/2,\,1,\,-1/2,\,-1\Bigr)\,+
\,C_2\,G_{04}^{10}\Bigl( z\,|1,\,1/2,\,-1/2,\,-1\Bigr)\,.
\label{solutiong}
\end{eqnarray}
We have also conditions
\begin{eqnarray}
& &1\,+\,8\int_0^{z_0}\,F(z)\,dz\,=\,
\frac{87\,g\,\sqrt{N}}{32\,\pi}\,\int_0^{z_0}F_0(z)\,\frac{dz}{\sqrt{z}}\,;\label{g}\\
& &F(z_0)\,=\,0\,;\label{pht1}
\end{eqnarray}
and boundary conditions analogous to~(\ref{bc}). The last
condition~(\ref{pht1}) means smooth transition from the non-trivial
solution to trivial one $G\,=\,0$. Knowing form~(\ref{solutiong}) of
a solution we calculate both sides of relation~(\ref{eqFg}) in two
different points in interval $0\,<\,z\,<\,z_0$ and having four
equations for four parameters solve the set. With $N\,=\,2$ we obtain 
the following solution, which we use to describe the electro-weak case
\begin{equation}
g(z_0)\,=\,-\,0.43014\,;\quad z_0\,=\,205.42535\,;\quad
C_1\,=\,0.003687\,; \quad C_2\,=\,0.005821\,.\label{gY}
\end{equation}
We would draw attention to the fixed value of parameter $z_0$. The solution
exists only for this value~(\ref{gY}) and it plays the role of eigenvalue.
As a matter of fact from the beginning the existence of such eigenvalue is
by no means evident.

Note that there is also solution with a smaller value of $z_0$ and large
positive $g(z_0)$, which with $N = 3$ 
presumably corresponds to strong interaction. This solution
is similar to that considered in work~\cite{Arb07} and it will be studied
elsewhere.

We consider the neglected terms of equation~(\ref{eqF}) as perturbations to
be taken into account in forthcoming studies.

\section{Running EW coupling}

We use Schwinger-Dyson equation for $W$-boson polarization operator to
obtain a contribution of additional effective vertex to the running EW
coupling constant $\alpha_{ew}$. The corresponding diagram is presented at
Fig. 2 a. Due to this vertex being gauge invariant,
there is no contribution of ghost fields. So the contribution under
discussion reads
\begin{equation}
\Delta \Pi_{\mu \nu}(x)\,=\,\frac{g\,G\,N}{2\,(2\,\pi)^4}\int \frac{\Gamma^0_
{\mu \rho \sigma}(p,-q-\frac{p}{2},q-\frac{p}{2})\,\Gamma^{eff}_{\nu \rho
\sigma}(-p,q+\frac{p}{2},-q+\frac{p}{2})\,F(q^2+\frac{3 p^2}{4})
\,dq}
{(q^2+p^2/4)^2\,-\,(p q)^2}\,;\label{SD}
\end{equation}
where $\Gamma^0_{\mu \rho \sigma}(p,q,k)\,=\,g_{\mu\rho}(p_\sigma-q_\sigma)+
g_{\rho\sigma}(q_\mu-k_\mu)+g_{\sigma\mu}(k_\rho-p_\rho)$ and $
\Gamma^{eff}_{\mu \rho \sigma}(p,q,k) $ is the Lorentz structure of
effective vertex~(\ref{vertex}).

After angular integrations we have
\begin{eqnarray}
& &\Delta \Pi_{\mu \nu}(x)\,=\,(g_{\mu \nu}\,p^2 -
p_\mu p_\nu)\,\Pi(x)\,;\quad x\,=\,p^2\,;\quad y'\,=\,q^2+\frac{3 x}{4}\,;
\nonumber\\
& &\Pi(x)\,=\,-\, \frac{g\,G\,N}{32\,\pi^2}\Biggl(\frac{1}{x^2}
\int_{3 x/4}^x
\frac{F(y') dy'}{y'-x/2}\,\biggl(16\frac{y'^3}{x^2}-48\frac{y'^2}{x}+45 y-
\frac{27}{2}x\biggr)\,+\label{DF}\\
& &+\,\int_x^Y \frac{F(y') dy'}{y'-x/2}\,\biggl(-\,3 y'\,+\,\frac{5}{2}\,x
\biggr)\Biggr)\,.\nonumber
\end{eqnarray}
Here coupling constant $g$ corresponds to $g(Y)$. We calculate integrals in~(\ref{DF}) with  substitution of solution~(\ref{solution}, \ref{gY}) numerically.

So we have modified one-loop expression for $\alpha_{ew}(p^2)$
\begin{equation}
\alpha_{ew}(x)\,=\,\frac{6\,\pi\,\alpha_{ew}(x'_0)}{6\,\pi\,+\,5\,\alpha_{ew}(x_0')
\ln(x/x_0')\,+\,6\,\pi\,\Pi(x)}\,; \quad x=p^2\,; \label{al1}
\end{equation}
where $x_0'$ means a normalization point such that $\Pi(x'_0)=0$. We normalize the running coupling
by condition
\begin{equation}
\alpha_{ew}(x_0)\,=\,\frac{g(Y)^2}{4\,\pi}\,;\label{nalpha}
\end{equation}
where
Coupling constant $g$ entering in expression
~(\ref{DF}) is just corresponding to this normalization point.
However in expression~(\ref{al1}) $x_0'$ does not coincide with $x_0$,
because polarization operator~(\ref{DF}) does not vanish at this point.
It does vanish at $ x'_0= 4/3\,x_0$. So we have to renormalize expression
~(\ref{al1}) and obtain
\begin{equation}
\alpha_{ew}(x'_0)\,=\,\alpha_{ew}(x_0)\frac{6 \pi(1+\Pi(x_0))}{6\pi+5\,\alpha_{ew}(x_0)\ln(4/3)}\,;  \label{al0}
\end{equation}
Using expressions~(\ref{al1}, \ref{nalpha}, \ref{al0}) we calculate behaviour of
$\alpha_{ew}(x)$ down to values of $x=p^2$ being by order of magnitude of $M_W$.

On this stage we have to get an information on our $G$. 
In the next section we define self-consistent value of this parameter so
achieve unique definition of $\alpha_{ew}(p^2)$.

\section{Interaction with Higgs and muon g-2 anomaly}

Let us consider a contribution of effective interaction
in~(\ref{Lint}) to $g-2$ anomaly. In the approach under
consideration this problem is connected with new contributions to
interaction of $W$-bosons and the Higgs particle. For the moment we
do not consider the scheme for electro-weak symmetry breaking in our
approach. For comparison with the actual physics we shall use the ready
phenomenology according to which the $W\,W\,H$ interaction
corresponds to the following vertex
\begin{equation}
\imath\,g_{\mu \nu}\,\delta_{a b}\,g\, M_W\,.\label{wwh0}
\end{equation}
From this moment we assume that $M_W$ is known. 
Additional contribution to vertex~(\ref{wwh0}) is provided by our effective
interaction due to diagram presented at Fig. 3. Substituting in the
first approximation $F_0$~(\ref{solution}, \ref{z0C}) we
obtain the following additional gauge invariant contribution
\begin{equation}
\imath\,2\,\sqrt{2}\,G\,g\,M_W\,\delta_{a b}\,(g_{\mu
\nu}\,(q\,k)\,-\,q_\nu\,k_\mu)\,F_H(x);\label{wwh1}
\end{equation}
where $F_H(x),\, F_H(0) = 1$ is a form-factor, which one calculate
from diagram Fig. 3. Calculation of an additional contribution to
anomalous magnetic moment of muon due to vertex~(\ref{wwh1}) needs
knowledge of a form-factor of $W\,W\,H$ interaction. Let us
formulate the equation for this form-factor, which is presented in
diagrams at Fig. 4, where inhomogeneous part is just
expression~(\ref{wwh1}), where we take the first two terms of
expansion of form-factor $F_H(x)$. Now the equation reads
\begin{eqnarray}
& &\Phi(x)\,=\,\frac{2\,G\,g\,M_W\,\sqrt{2}}{H}\,+\,\frac{g\,M_W\,G^2\,x}{16 \pi^2\,H}
\biggl(\ln\,z\,+\,4\,\gamma+4 \ln 2\,-\,\frac{16}{3}\,+\nonumber\\
& &+\,\frac{1}{2}\,G_{15}^{31}\Bigl( z_0 |^0_{0,\,0,\,1/2,\,-1/2,\,-1}\Bigr)\biggr) +
\,\beta \Biggr(\frac{1}{2} \int_0^Y \Phi(y)\,dy +
\frac{1}{12\,x^2}\int_0^x y^2\,\Phi(y)\,dy -\nonumber\\
& &-\,\frac{1}{6\,x}\int_0^x\,y\,\Phi(y)\,dy\,-\,\frac{x}{6}\,\int_x^Y\frac{\Phi(y)}{y}\,dy\,+\,\frac{x^2}{12}\,
\int_x^Y\frac{\Phi(y)}{y^2}\,dy\Biggr)\,;\label{EQH}\\
& &\beta\,=\,\frac{h^2}{16\,\pi^2}\,; \quad h\,=\,\frac{4\,\sqrt{2}\,G\,g\,M_W}{2\,-\,\beta\,\int_0^Y\,\Phi(y)\,dy}\,
\,; \quad \Phi(0)\,=\,1\,.\nonumber
\end{eqnarray}
Here $h$ is a resulting constant of $W W H$ interaction and $\Phi(x)$ is the
form-factor of the interaction with momentum $q \to 0$. Upper limit
of integration $Y$ has to be the same as in~(\ref{eqF}), because our
result in studying non-trivial solution of~(\ref{eqF}) demands
$G\,=\,0$ for $x\,>\,Y$. The same $G$ enters into
expression~(\ref{wwh1}) so that for $x\,>\,Y$ we also demand
$\Phi(x)$ to vanish. Thus we have also condition
\begin{equation}
\Phi(Y)\,=\,0\,.\label{phi0}
\end{equation}
This condition defines relation between variable $z$ of the previous section and dimensionless variable
$u$, which is peculiar to Eq.~(\ref{EQH})
\begin{eqnarray}
& &u = \beta\,x\,=\,\frac{8 \sqrt{2}\,g^2\,M_W^2\,G}{\pi\,\eta^2}\,\sqrt{z}=
\,\xi\,\sqrt{z}\,;\label{ksi}\\
& &\eta\,=\,1\,-\,\frac{1}{2}\,\int_0^{u_0}\,\Phi(u)\,du\,;\quad
u_0\,=\,\xi\,\sqrt{z_0}\,.\nonumber
\end{eqnarray}
Performing substitutions~(\ref{ksi}) into~(\ref{EQH}) we obtain the following
equation
\begin{eqnarray}
& &\Phi(u)\,=\,1\,+\,\frac{\eta\,u}{2\,\pi\,\xi}
\biggl(2\ln\,u\,-\,2\ln\,\xi\,+4\,\gamma\,+\,4 \ln 2\,-\,\frac{16}{3}\,+\nonumber\\
& &+\,\frac{1}{2}\,G_{15}^{31}\Bigl( z_0\, |^0_{0,\,0,\,1/2,\,-1/2,\,-1}\Bigr)\biggr)
\, +\,
\frac{1}{12\,u^2}\int_0^u t^2\,\Phi(t)\,dt -\label{EQu}\\
& &-\,\frac{1}{6\,u}\int_0^u\,t\,\Phi(t)\,dt\,-\,\frac{u}{6}\,\int_u^{u_0}\frac{\Phi(t)}{t}\,dt\,+\,\frac{u^2}{12}\,
\int_u^{u_0}\frac{\Phi(t)}{t^2}\,dt\,;\nonumber
\end{eqnarray}
We look for a solution of~(\ref{EQu}) with condition~(\ref{phi0}) in the
following form
\begin{eqnarray}
& &\Phi(u)\,=\,2\,G^{31}_{15}\biggl(\,u\,|^0_{0,1,2,-2,-1}\biggr)\,-
\,\frac{12\,\eta}{\pi\,\xi}\,G^{31}_{15}\biggl(\,u\,|^1_{1,1,2,-2,-1}\biggr)\,
+\nonumber\\
& &+\,\,C^\phi_2\,\biggl(G^{20}_{04}(-\,u\,|1,2,-2,-1)\,-
\,\imath\,\pi\,G^{10}_{04}(\,u\,|2,-2,-1,1)\biggr)\,+\label{solphi}\\
& &+\,C^\phi_1\,G^{10}_{04}(\,u\,|2,-2,-1,1)\,;\quad \Phi(u_0)\,=\,0\,.\nonumber
\end{eqnarray}
Now we substitute~(\ref{solphi}) into~(\ref{EQu})
and obtain the following solution for parameters
\begin{equation}
C^\phi_1\,=\,0.721216\,;\quad C^\phi_2\,=\,-\,4.027240\,;\quad \eta\,=\,
-\,0.077332\,;\quad \xi\,=\,0.2872314\,.\label{solcphi}
\end{equation}
Relation~(\ref{ksi}) allows to define coupling constant $G$ of
the three-boson effective interaction
\begin{equation}
G\,=\,\frac{\Lambda}{M_W^2}\,;\quad \Lambda\,=\,0.010312\,.\label{Lambda}
\end{equation}
Note, that $g$ in relation~(\ref{ksi}) is just $g(z_0)$ from solution~(\ref{gY}). 
With this result we completely define expression for $\alpha_{ew}$ of the previous
section. So substituting~(\ref{Lambda}) into relations~(\ref{al1}, \ref{nalpha}, \ref{al0}) we obtain
\begin{equation}
\alpha_{ew}(M_W^2)\,=\,0.0374\,;\label{alphaew}
\end{equation}
what is only $10\%$ larger than well-known value
\begin{equation}
\alpha_{ew}^{exp}(M_W^2)\,=\,\frac{\alpha(M_W)}{\sin^2_W}\,=\,0.0337\,.
\label{alphaewexp}
\end{equation}
We consider this result as strong confirmation of the approach. As a
matter of fact the accuracy of the present approach was estimated to
be just $(10\,-\,15)\%$~\cite{Arb04, AVZ}.

From relations~(\ref{Lambda}, \ref{alphaew}) bearing in mind negative sign 
of $g$ we have
\begin{equation}
\lambda\,=\,\frac{\Lambda}{g(M_W^2)}\,=\,-\,0.0151\,;\label{lambda2}
\end{equation}
that evidently agrees with limitations~(\ref{lambda1}).

Now we have new effective interaction of Higgs with $W$ with
form-factor~(\ref{solphi}, \ref{solcphi}). Due to diagrams of Fig. 5
this interaction contributes to muon magnetic moment giving the
following additional term in $a\,=\,g-2$
\begin{eqnarray}
& &\Delta a\,=\,-\,\frac{\Lambda\,\sqrt{2}}{8\,\pi\,\eta}\Biggl(\frac{m_\mu}{M_W}
\Biggr)^2\,\int_0^{u_0} \frac{\alpha_{ew}(u)\,\Phi(u)\,u\,du}{(u+u_w)(u+u_h)};
\label{da}\\
& &u_w\,=\,\frac{h^2\,M_W^2}{16\,\pi^2}\,;\quad u_h\,=\,\frac{h^2\,M_H^2}{16\,\pi^2}
\,=\,u_w\,\frac{M_H^2}{M_W^2}\,;\nonumber
\end{eqnarray}
where $M_H$ is yet unknown mass of the Higgs particle.
Behaviour~(\ref{al1}) of $\alpha_{ew}(Q)$ is presented at Fig. 6.

We know everything but $M_H$ in expression~(\ref{da}) and e.g. for 
mass of Higgs $M_H\,=\,114\,GeV$ we obtain
\begin{equation}
\Delta a\,=\,3.34\cdot 10^{-9};\label{da1}
\end{equation}
that comfortably fits into error bars for well-known deviation~\cite{g-2, g-2t1, g-2t2}
\begin{equation}
\Delta a\,=\,(3.02\,\pm\,0.88)\cdot 10^{-9}\,.\label{daexp}
\end{equation}
With  $M_H$ growing $\Delta a$~(\ref{da}) slowly decreases  inside the error bars down to
$2.67\cdot 10^{-9}$ for $M_H = 300\,GeV$. Thus we can state, that our result agrees 
with experiment~(\ref{daexp}) for values $150\,GeV\,<\,M_H\,<\,300\,GeV$, which we 
shall discuss in the next section in connection with experimental implications. 

Contribution~(\ref{da}) to electron $g-2$ is four orders of magnitude smaller and 
so it is far below experimental accuracy $\pm\,4\cdot 10^{-12}$.

\section{Experimental implications}
New interaction of $H$ with $W$-s
\begin{equation}
L_{HWW}\,=\,\frac{h}{2}\,W^a_{\mu \nu}\,W^a_{\mu \nu}\,H.\label{inthww1}
\end{equation}
leads to changes in usual branching ratios for $H$ decays. We use here the well-known expression for $W^0$  mixed state with physical value for $\sin^2\theta_W$. There are unusually significant channels 
$H \to \gamma\,\gamma$ and $H \to \gamma\,Z$. Therefore there are additional restrictions from existing experiments. Recent data from Tevatron on search for 
Higgs particle in $\gamma\,\gamma$ channel~\cite{ggd0} exclude Higgs particle with 
interaction~(\ref{inthww1}) for $M_H\,<\,150\,GeV$. Thus in the framework of our approach we consider only
\begin{equation}
M_H\,>\,150\,GeV\,.\label{restrh}
\end{equation}
We calculate cross-sections of Higgs production at Tevatron with $\sqrt{s} = 1960\,GeV$ and branching ratios of its decays for three values of the mass: $175,\,200,\,225\,GeV$. Results are presented in Table 1. 
We see, that presumably it could be possible to study our predictions with the existing Tevatron facilities in channels $\gamma\,\gamma,\, \gamma\,Z,\, Z\,Z,\,W^+\,W^-$. A choice of 
the most promising channel depends on efficiency of registration. 

We present also in Table 2 predictions for forthcoming LHC search for Higgs particle. 

For the sake of further experimental implications we present behaviour of
$\alpha_{ew}(Q)$ at Fig. 6, where also the usual perturbative
dependence of $\alpha_{ew}(Q)$ is shown. The main difference of the
solution from perturbative description consists in existence of
triple gauge-boson interaction~(\ref{effint}) with
form-factor~(\ref{solutiong}), (\ref{gY}). The  dependence of this
form-factor on $Q$ is presented at Fig. 7. From these pictures one
sees that the difference of physical effects from the perturbative 
ones might occur for $TeV$ range of energy. 

First of all we calculate total cross-section of reaction
$e^+\,e^-\,\to\,W^+\,W^-$ for our solution. The result is presented
at Fig. 8 for c.m.r.f. total energy $0.4\,TeV < Q <9\,TeV$. One sees, that the 
difference of the non-perturbative solution, which corresponds to the upper
curve, from the perturbative one (lower curve) starts around
$1.5\,TeV$ and becomes maximal at $5\,-\,6\,TeV$. Really for small
$Q$ contribution of the three-boson interaction is small. Then the
contribution increases with $Q$ increasing, but for much larger $Q$
the decrease of the form-factor (see Fig. 7) in line with decreasing
$\alpha_{ew}(Q)$ lead to fast decrease of the effect.

Let us consider also reaction $e^+e^-\to Z\,Z$. The behavior of the cross-section is
presented at Fig. 9. Here the non-perturbative curve is below the perturbative one. The effect seems significant, but cross-section of the reaction in $TeV$ region
is small contrary to $e^+\,e^-\,\to\,W^+\,W^-$.

Implication of our results for reactions at LHC
\begin{equation}
p\,+\,p\,\to\,W^+\,+\,W^-\,+X\,;\quad p\,+\,p\,\to\,W^+\,+\,Z\,+X\,;\quad p\,+\,p\,\to\,Z\,+\,Z\,+X\,;
\label{W+W-}
\end{equation}
needs special extensive study. One could expect here significant effects for 
high invariant masses of boson pairs.

For calculations of this section CompHEP package~\cite{comphep} was used. 

\section{Conclusion}

To conclude the author would emphasize, that albeit we discuss quite
unusual effects, we do not deal with something beyond the Standard
Model. We are just in the framework of the Standard Model. What
makes difference with usual results is {\bf non-trivial solution} of
compensation equation. There is of course also {\bf trivial
perturbative solution}. Which of the solutions is realized is to be
defined by a stability condition. In view of this it would be desirable to estimate
contribution of our solution to vacuum energy density. For example 
in our case we have non-zero non-perturbative boson
condensate, 
which in diagram form is presented at Fig. 2 b. It is positive, that
means negative contribution to vacuum energy density. So with
account of only this fact we could state that non-trivial solution is
stable and therefore it really exists. Of course real situation is much more 
difficult and there are
other contributions to the vacuum energy density, which have to be
taken into account. First of all the contribution of symmetry
breaking mechanism, e.g. of the Higgs one, is without doubt quite
important. For the moment we would say, that the problem of
stability will be considered in forthcoming studies in more details.

With the present results we would draw attention to two important
achievements provided by the non-trivial non-perturbative solution.
The first one is unique determination of gauge electro-weak coupling
constant~(\ref{alphaew}) in close agreement with experimental
value~(\ref{alphaewexp}). At this point we would emphasize, that the
existence of a non-trivial solution itself always leads to
additional conditions for parameters of a problem under study. So
the result on the coupling constant is by no means surprising. The
second important point consists in agreement of our calculation 
for additional contribution to muon anomalous magnetic
moment~(\ref{da1}) with experimental number~(\ref{daexp}). So this effect does not 
need a hypothetical exit beyond the Standard Model. These two
achievements strengthen the confidence in the correctness of
applicability of Bogoliubov compensation approach to the principal
problems of elementary particles theory.

Recent Tevatron data impose on mass of Higgs particle restriction $M_H\,>\,150\,GeV$. 
It seems, that Tevatron facilities allow to check predictions of the work for $M_H$ 
starting from this value up to value $\simeq 250\,GeV$. 

We have also pointed to the possibility of verification of the present results in 
forthcoming experiments at LHC. The most promising processes are again Higgs production and presumably pair production 
of weak bosons $W^+\,W^-$ and $W^+\,Z$ with high invariant masses. 

\section*{Acknowledgments}

The author express gratitude to E.E. Boos and V.I Savrin for valuable discussions.

\newpage
\begin{center}
{\bf Table captions}
\end{center}
\bigskip
\bigskip
Table 1. Predictions for Tevatron search for Higgs particle, $\sqrt{s}\,=\,1.96\,TeV$.\\
\\
Table 2. Predictions for LHC search for Higgs particle, $\sqrt{s}\,=\,14\,TeV$.

\newpage
\begin{center}
{\bf Figure captions}
\end{center}
\bigskip
\bigskip
Fig. 1. Diagram representation of the compensation
equation. Black spot corresponds to anomalous three-boson
vertex with a form-factor. Empty circles correspond to point-like anomalous
three-boson and four-boson vertices. Simple point corresponds to usual gauge  vertex.
Incoming momenta are denoted by the corresponding external lines.\\
\\
Fig. 2. Loop contribution to boson polarization operator. Simple point
corresponds to the perturbative vertex (a). Diagram corresponding to calculation
of boson condensate; cross on the inside line describes
"vertex" $W_{\mu \nu}^a\,W_{\mu \nu}^a$ (b).\\
\\
Fig. 3. Diagram for the first step in calculation of $H\,W\,W$ vertex. Double line represents Higgs particle.
Simple point represent usual $H\,W\,W$ vertex.\\
\\
Fig. 4. Diagram representation of the equation for $H\,W\,W$ form-factor. Double circle with black internal one corresponds to anomalous  $H\,W\,W$
vertex with a form-factor. The same with empty internal circle correspond to point-like anomalous  $H\,W\,W$ vertex. Single circle corresponds to $H\,W\,W$ vertex
calculated according diagram at Fig. 3.\\
\\
Fig. 5. Diagrams for new contribution to muon magnetic moment. Dotted line represents
muon (or other spin one-half particle), the line going up from the full vertex describes a photon.\\
\\
Fig. 6. Behaviour of modified $\alpha_{ew}(Q)$ for $0 < Q < 40\,TeV$
with  $\alpha_{ew}(M_W) = 0.0374$. The upper line corresponds to usual electro-weak
coupling with the same normalization. \\
\\
Fig. 7. Behaviour of form-factor $F(Q)$, $0 < Q < 35.74\,TeV$. For $Q >35.74\,TeV$
$F(Q)\,=\,0$.\\
\\
Fig. 8. Total cross-section  $\sigma(Q)$ of reaction $e^+\,e^-\,\to\,W^+\,W^-$
for~$0.6 \, TeV  <  Q < 9\,TeV$. The lower line corresponds to usual electro-weak
coupling.\\
\\
Fig. 9. Total cross-section  $\sigma(Q)$ of reaction $e^+\,e^-\,\to\,Z\,Z$
for~$0.6 \, TeV  <  Q < 7\,TeV$. The upper line corresponds to usual electro-weak
coupling.

\newpage
\begin{center}
{\bf Table 1.} \\
\bigskip
\bigskip
\large{
\begin{tabular}{|c|c|c|c|}
\hline
$M_H\,GeV$ & 175 & 200 & 225 \\
\hline
$\Gamma_H\,GeV$ & 0.287 & 0.789 & 1.714  \\
\hline
$\sigma_t(p\,\bar p \to H + X)\,pb$ & 0.885 & 0.586 & 0.403 \\
\hline
$BR(H \to \gamma\,\gamma)$ \% & 7.41 & 4.02 & 2.63  \\
\hline
$BR(H \to \gamma\,Z)$ \% & 25.60 & 17.84 & 13.73  \\
\hline 
$BR(H \to Z\,Z)$ \% & 0  & 12.65 & 17.12 \\
\hline
$BR(H \to W^+\,W^-)$ \% & 64.38 &  64.40 & 65.95  \\
\hline
$BR(H \to b\,\bar b)$ \% & 2.61 & 1.09 & 0.56 \\
\hline 
$\sigma_t\,BR(\gamma\,\gamma)\,pb$ & 0.066 & 0.024 & 0.011 \\
\hline
$\sigma_t\,BR(\gamma\,Z)\,pb$ & 0.227 & 0.105 & 0.055 \\
\hline 
$\sigma_t\,BR(Z\,Z)\,pb$ & 0 & 0.074 & 0.069 \\
\hline
$\sigma_t\,BR(W^+\,W^-)\,pb$  & 0.570 & 0.377 & 0.266 \\
\hline
$\sigma_t\,BR(b\,\bar b)\,pb$  & 0.023 & 0.006 & 0.002  \\
\hline
\end{tabular}
}
\end{center}
\newpage
\begin{center}
{\bf Table 2.}\\
\bigskip
{\large
\begin{tabular}{|c|c|c|c|c|c|}
\hline
$M_H\,GeV$ & 175 & 225 & 275 & 325 & 375\\
\hline
$\Gamma_H\,GeV$ & 0.287 & 1.714 & 2.474 & 12.306 & 23.808\\
\hline
$\sigma_t(p\,p \to H + X)\,pb$ & 6.54 & 4.48 & 3.44 & 2.60 & 2.11 \\
\hline
$\sigma_t\,BR(\gamma\,\gamma)\,pb$ & 0.485 & 0.118 & 0.052 & 0.029 & 0.019\\
\hline
$\sigma_t\,BR(\gamma\,Z)\,pb$ & 1.674 & 0.615 & 0.326 & 0.201 & 0.138 \\
\hline
$\sigma_t\,BR(Z\,Z)\,pb$ & 0 & 0.766 & 0.827 & 0.720 & 0.597 \\
\hline 
$\sigma_t\,BR(W^+\,W^-)\,pb$ & 4.210  & 2.952 & 2.227 & 1.650 & 1.258 \\
\hline
$\sigma_t\,BR(b\,\bar b)\,pb$ & 0.171 &  0.025 & 0.007 & 0.003 & 0.001\\
\hline
$\sigma_t\,BR(t\,\bar t)\,pb$ & 0 &  0 & 0 & 0 & 0.095\\
\hline
\end{tabular}
}
\end{center}

\newpage
\begin{picture}(160,105)
{\thicklines
\put(5,90.5){\line(-3,2){10}}
\put(5,90.5){\line(-3,-2){10}}
\put(5,90.5){\circle*{3}}}
\put(5,90.5){\line(1,0){13}}
\put(-5,100.5){p}
\put(-5,80.5){-p}
\put(10,92.5){0}
\put(23.5,90){+}
{\thicklines
\put(52.5,90.5){\line(-3,2){15}}
%\put(52.5,60.5){\oval(20,10)[t]}
\put(52.5,90.5){\line(-3,-2){15}}
\put(37.5,102.5){p}
\put(37.5,77.5){-p}
\put(57.5,92.5){0}
\put(52.5,90.5){\circle*{3}}
\put(42.5,83.5){\line(0,1){13.4}}
\put(42.5,83.8){\circle{3}}
\put(42.5,97){\circle{3}}}
%\put(2.5,80.5)
%{\line(1,-1){5}}
\put(52.5,90.5){\line(1,0){13}}}
\put(80,90){+}
{\thicklines
\put(102.5,90.5){\line(-3,2){10}}
\put(102.5,90.5){\line(-3,-2){10}}
\put(102.5,90.5){\circle{3}}
\put(112.5,90.5){\oval(20,10)}
\put(122.5,90.5){\line(1,0){13}}
\put(122.5,90.5){\circle*{3}}
\put(92,100.5){p}
\put(92,80.5){-p}
\put(127.5,92.5){0}
\put(80,90){+}
{\thicklines
\put(102.5,90.5){\line(-3,2){10}}
\put(102.5,90.5){\line(-3,-2){10}}
\put(102.5,90.5){\circle{3}}
\put(112.5,90.5){\oval(20,10)}
\put(122.5,90.5){\line(1,0){13}}
\put(122.5,90.5){\circle*{3}}}
\put(0,50){+}
{\thicklines
\put(32.5,60.5){\line(-3,2){10}}
\put(32.5,40.5){\line(-3,-2){10}}
\put(32.5,50.5){\oval(10,20)}
\put(32.5,40.5){\line(1,0){13}}
\put(32.5,60.5){\circle*{3}}
\put(32.5,40.5){\circle{3}}
\put(22.5,70.5){p}
\put(22.5,30){-p}
\put(40.5,42.5){0}}
\put(60,50){+}
{\thicklines
\put(92.5,40.5){\line(-3,-2){10}}
\put(92.5,60.5){\line(-3,2){10}}
\put(92.5,50.5){\oval(10,20)}
\put(92.5,60.5){\line(1,0){13}}
\put(92.5,40.5){\circle*{3}}
\put(92.5,60.5){\circle{3}}
\put(82.5,70.5){p}
\put(82.5,30){-p}
\put(100.5,62.5){0}}
%\put(23.5,50){+}
\put(0,10){+}
{\thicklines
\put(22.5,10.5){\line(-3,2){15}}
%\put(52.5,60.5){\oval(20,10)[t]}
\put(22.5,10.5){\line(-3,-2){15}}
\put(7.5,22.5){p}
\put(7.5,-3.5){-p}
\put(27.5,12.5){0}
\put(22.5,10.5){\circle*{3}}
\put(12.5,3.5){\line(0,1){13.4}}
\put(12.5,3.8){\circle{3}}
%\put(12.5,16.8){\circle{3}}
\put(22.5,10.5){\line(1,0){13}}}
\put(40,10){+}
{\thicklines
\put(62.5,10.5){\line(-3,2){15}}
%\put(52.5,60.5){\oval(20,10)[t]}
\put(62.5,10.5){\line(-3,-2){15}}
\put(47.5,22.5){p}
\put(47.5,-3.5){-p}
\put(67.5,12.5){0}
\put(62.5,10.5){\circle*{3}}
\put(52.5,3.5){\line(0,1){13.4}}
%\put(52.5,3.8){\circle{3}}
\put(52.5,17){\circle{3}}
\put(62.5,10.5){\line(1,0){13}}}
\put(80,10){+}
{\thicklines
\put(102.5,10.5){\line(-3,2){15}}
%\put(52.5,60.5){\oval(20,10)[t]}
\put(102.5,10.5){\line(-3,-2){15}}
\put(87.5,22.5){p}
\put(87.5,-3.5){-p}
\put(107.5,12.5){0}
%\put(102.5,10.5){\circle*{3}}
\put(92.5,3.5){\line(0,1){13}}
\put(92.5,3.8){\circle{3}}
\put(92.5,17){\circle*{3}}
\put(102.5,10.5){\line(1,0){13}}}
\put(120,10){=}
\put(130,10){0}
\end{picture}

\bigskip
\bigskip
\bigskip
\bigskip
\bigskip

\begin{center}
Fig. 1.
\end{center}
\newpage
\begin{picture}(160,30)
{\thicklines
\put(32.5,20.5){\line(-1,0){15}}
\put(42.5,20.5){\oval(20,10)}
\put(52.5,20.5){\line(1,0){15}}
\put(25.5,22.5){p}
\put(52.5,20.5){\circle*{3}}
%\put(82.5,53.5){\line(0,1){13}}
\put(60.5,22.5){-p}}
\put(40,5){(a)}
{\thicklines
\put(92.5,20.5){\line(1,0){20}}
\put(102.5,20.5){\oval(20,10)}
\put(101.3,19.4){x}
%\put(55.5,22.5){p}
\put(92.5,20.5){\circle*{3}}
\put(112.5,20.5){\circle*{3}}}
%\put(82.5,53.5){\line(0,1){13}}}
%\put(90.5,22.5){-p}}
\put(100,5){(b)}
\end{picture}
\\
\\

\begin{center}
Fig. 2.
\end{center}

\newpage
\begin{picture}(160,30)
{\thicklines
\put(86,20){q  $\nu$}
\put(57,35){k  $\mu$}
\put(82.5,17){\line(-3,2){30}}
%\put(52.5,60.5){\oval(20,10)[t]}
\put(82.5,17){\line(-3,-2){20}}
\put(62.5,4.6){\line(-3,-2){10}}
\put(62.5,3.4){\line(-3,-2){10}}
%\put(47.5,22.5){p}
%\put(47.5,-3.5){-p}
%\put(67.5,12.5){0}
\put(82.5,17){\circle*{3}}
\put(62.5,3.5){\line(0,1){26.8}}
%\put(52.5,3.8){\circle{3}}
\put(62.5,30){\circle{3}}
\put(82.5,17){\line(1,0){13}}}
\end{picture}
\\
\\
\\
\\
\\

\begin{center}
Fig. 3.
\end{center}

\newpage
\begin{picture}(160,30)
{\thicklines
\put(12.5,17){\line(-3,2){20}}
%\put(32.5,16){\line(-3,2){20}}
%\put(52.5,60.5){\oval(20,10)[t]}
\put(12.5,17){\line(-3,-2){20}}
\put(12.5,18){\line(-3,-2){20}}
%\put(12.5,4.6){\line(-3,-2){10}}
%\put(12.5,3.4){\line(-3,-2){10}}
%\put(47.5,22.5){p}
%\put(47.5,-3.5){-p}
%\put(67.5,12.5){0}
\put(12.5,17){\circle*{3}}
\put(12.5,17){\circle{5}}
%\put(12.5,3.5){\line(0,1){26.8}}
%\put(52.5,3.8){\circle{3}}
%\put(12.5,30){\circle{3}}
%\put(12.5,4){\circle{3}}
\put(12.5,17){\line(1,0){13}}
\put(18,20){q}
\put(2.5,26){k}
\put(32.5,16){=}

\put(62.5,17){\line(-3,2){15}}
%\put(32.5,16){\line(-3,2){20}}
%\put(52.5,60.5){\oval(20,10)[t]}
\put(62.5,17){\line(-3,-2){15}}
\put(62,18){\line(-3,-2){15}}
%\put(12.5,4.6){\line(-3,-2){10}}
%\put(12.5,3.4){\line(-3,-2){10}}
%\put(47.5,22.5){p}
%\put(47.5,-3.5){-p}
%\put(67.5,12.5){0}
%\put(62.5,17){\circle*{3}}
\put(62.5,17){\circle{5}}
%\put(12.5,3.5){\line(0,1){26.8}}
%\put(52.5,3.8){\circle{3}}
%\put(12.5,30){\circle{3}}
%\put(12.5,4){\circle{3}}
\put(62.5,17){\line(1,0){13}}
\put(82.5,16){+}
\put(68,20){q}
\put(55,24){k}
\put(128,20){q}
\put(98,35){k}
\put(122.5,17){\line(-3,2){30}}
\put(122.5,16){\line(-3,2){20}}
%\put(52.5,60.5){\oval(20,10)[t]}
\put(122.5,17){\line(-3,-2){20}}
\put(102.5,4.6){\line(-3,-2){10}}
\put(102.5,3.4){\line(-3,-2){10}}
%\put(47.5,22.5){p}
%\put(47.5,-3.5){-p}
%\put(67.5,12.5){0}
\put(122.5,17){\circle*{3}}
\put(122.5,17){\circle{5}}
\put(102.5,3.5){\line(0,1){26.8}}
%\put(52.5,3.8){\circle{3}}
\put(102.5,30){\circle{3}}
\put(102.5,30){\circle{5}}
\put(102.5,4){\circle{3}}
\put(102.5,4){\circle{5}}
\put(122.5,17){\line(1,0){13}}}
\end{picture}
\\
\\
\\
\\
\\

\begin{center}
Fig. 4.
\end{center}

\newpage
\begin{picture}(160,30)
{\thicklines
\multiput(22,0)(3,0){15}%
{\line(1,0){2}}
\put(45,20){\line(0,1){10}}
\put(45,20){\line(2,-3){13.4}}
\put(44,20){\line(2,-3){13.4}}
\put(45,20){\line(-2,-3){13.4}}
\put(45,20){\circle*{3}}
\put(45,20){\circle{5}}
\multiput(87,0)(3,0){15}%
{\line(1,0){2}}
\put(110,20){\line(0,1){10}}
\put(110,20){\line(2,-3){13.4}}
\put(111,20){\line(-2,-3){13.4}}
\put(110,20){\line(-2,-3){13.4}}
\put(110,20){\circle*{3}}
\put(110,20){\circle{5}}
\put(77,10){+}
\put(45,-5){$\mu$}
\put(110,-5){$\mu$}

}
\end{picture}
\\
\\
\\
\\
\\

\begin{center}
Fig. 5.
\end{center}

\newpage

\begin{figure}[ht]
\begin{picture}(300,450)
\put(20,300){\epsfig{file=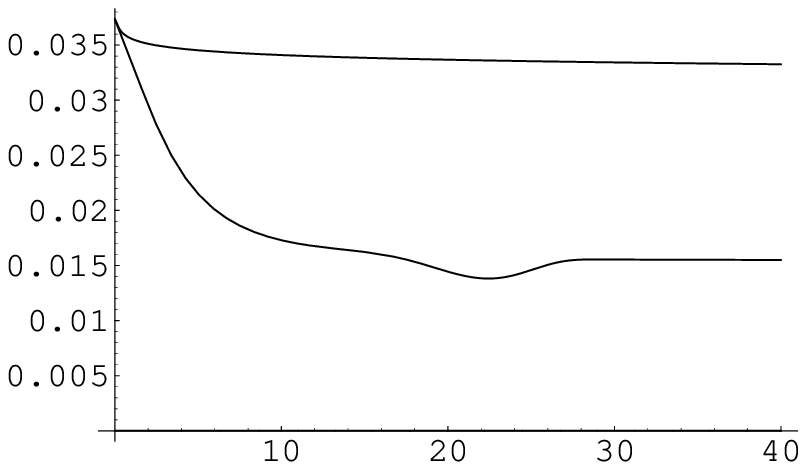,width=300pt,
height=450pt}}
                         % this instruction read file (s_chain.eps)
                         % end define size of picture in pt (width/height)
\put(60,270){Fig. 6}
\put(15,450){\Large {$\alpha_{ew}(Q)$}}
\put(130,309){\Large {$TeV$}}
\put(85,300){\Large {$Q$}}
\end{picture}

\caption{Behaviour of modified $\alpha_{ew}(Q)$, for $0 < Q < 1\,GeV$,
$\Lambda_{QCD}\,=\,0.2\,GeV$. The upper line corresponds to one-loop Shirkov -- Solovtsov
running coupling with  $\Lambda_{QCD}\,=\,0.25\,GeV$.
}

\label{}

\end{figure}
\newpage
\begin{figure}[ht]
\begin{picture}(300,450)
\put(20,300){\epsfig{file=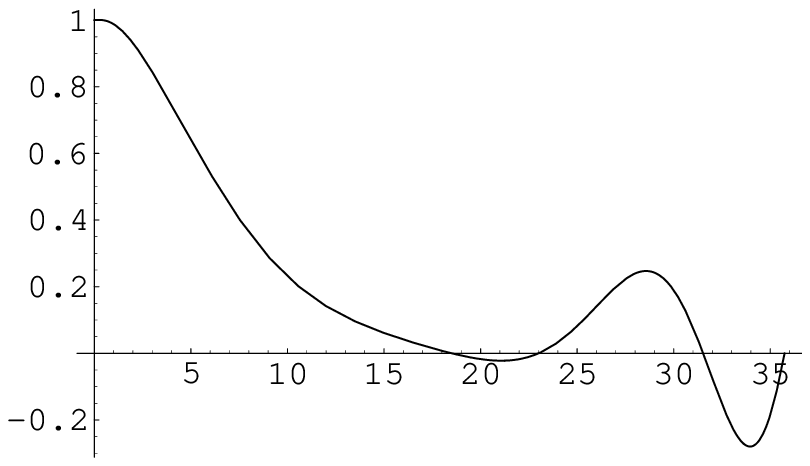,width=300pt, height=450pt}}
                         % this instruction read file (s_chain.eps)
                         % end define size of picture in pt (width/height)
\put(60,270){Fig. 7}
\put(10,440){\Large {$F(Q)$}}
\put(130,331){\Large {$TeV$}}
\put(75,322){\Large {$Q$}}
\end{picture}

\caption{Behaviour of modified $\alpha_{ew}(Q)$, for $0 < Q < 1\,GeV$,
$\Lambda_{QCD}\,=\,0.2\,GeV$. The upper line corresponds to one-loop Shirkov -- Solovtsov
running coupling with  $\Lambda_{QCD}\,=\,0.25\,GeV$.
}

\label{}

\end{figure}
\newpage
\begin{figure}[ht]
\begin{picture}(300,450)
\put(20,300){\epsfig{file=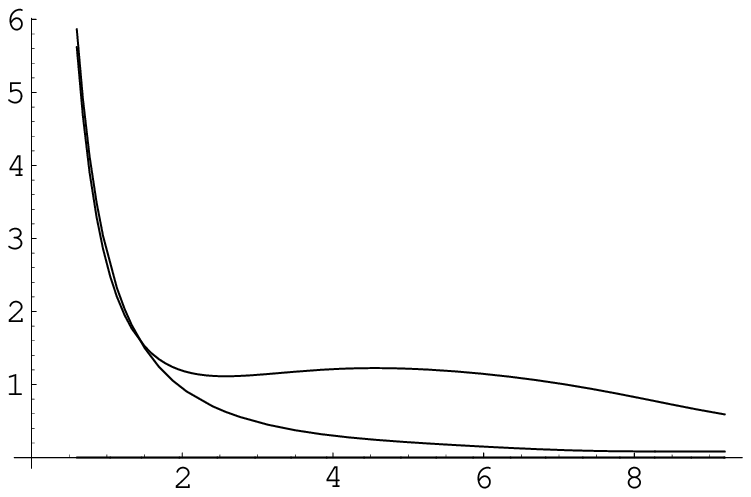,width=300pt, height=450pt}}
                         % this instruction read file (s_chain.eps)
                         % end define size of picture in pt (width/height)
\put(60,270){Fig. 8}
\put(5,445){\Large {$\sigma(Q)\,pb$}}
\put(70,400){\Large {$e^+e^- \to W^+W^-$}}
\put(115,304){\Large {$TeV$}}
\put(75,295){\Large {$Q$}}
\end{picture}

\caption{Behaviour of modified $\alpha_{ew}(Q)$, for $0 < Q < 1\,GeV$,
$\Lambda_{QCD}\,=\,0.2\,GeV$. The upper line corresponds to one-loop Shirkov -- Solovtsov
running coupling with  $\Lambda_{QCD}\,=\,0.25\,GeV$.
}

\label{}

\end{figure}
%\end{document}
\newpage
\begin{figure}[ht]
\begin{picture}(300,450)
\put(20,300){\epsfig{file=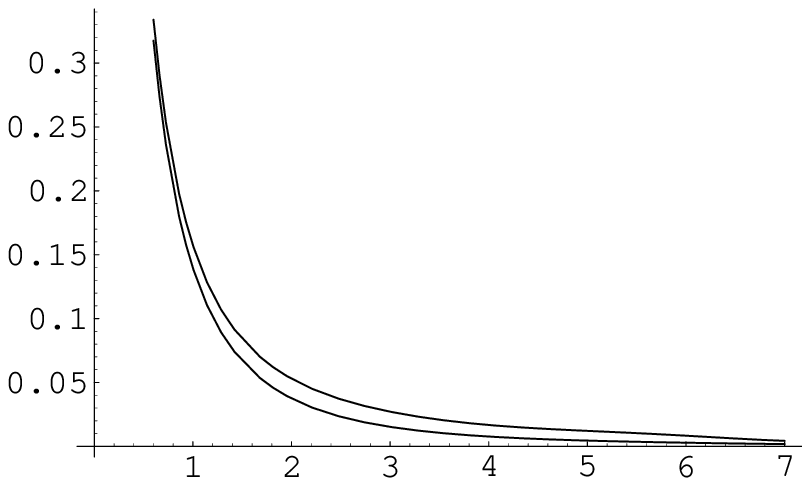,width=300pt, height=450pt}}
                         % this instruction read file (s_chain.eps)
                         % end define size of picture in pt (width/height)
\put(60,270){Fig. 9}
\put(5,445){\Large {$\sigma(Q)\,pb$}}
\put(70,400){\Large {$e^+e^- \to Z\,Z$}}
\put(128,307){\Large {$TeV$}}
\put(75,297){\Large {$Q$}}
\end{picture}

\label{}

\end{figure}


\begin{thebibliography}{**}
\bibitem{Arb04} B.A. Arbuzov, Theor. Math. Phys., {\bf 140}, 1205 (2004).
\bibitem{Arb05} B.A. Arbuzov, Phys. Atom. Nucl., {\bf 69}, 1588 (2006).
\bibitem{AVZ} B.A. Arbuzov, M.K. Volkov and I.V. Zaitsev, Int. Journ. Mod.
Phys. A, {\bf 21}, 5721 (2006).
\bibitem{Arb07} B.A. Arbuzov, Phys. Lett. B, {\bf 656}, 67 (2007).
\bibitem{AVZ2} B.A. Arbuzov, M.K. Volkov and I.V. Zaitsev, Int. Journ. Mod.
Phys. A (in press); arXiv:0809.4952 [hep-ph] (2008).
\bibitem{Bog1} N.N. Bogoliubov. Soviet Phys.-Uspekhi, {\bf 67}, 236 (1959).
\bibitem{Bog2} N.N. Bogoliubov. Physica Suppl., {\bf 26}, 1 (1960).
\bibitem{Bog} N.N. Bogoliubov, {\it Quasi-averages in problems of
statistical mechanics.} Preprint JINR D-781, (JINR, Dubna 1961).
\bibitem{Hag} K. Hagiwara, R.D. Peccei, D. Zeppenfeld and K. Hikasa, Nucl. Phys. B, {\bf 282}, 253 (1987).
\bibitem{Arb92} B.A. Arbuzov, Phys. Lett. B, {\bf 288}, 179 (1992).
\bibitem{EW} LEP Electro-weak Working Group, arXiv:hep-ex/0612034v2 (2006).
\bibitem{be} H. Bateman and A. Erd\'elyi, {\it Higher
transcendental functions. V. 1} (New York, Toronto, London: McGraw-Hill,
1953).
\bibitem{g-2} G.W. Bennett et al., Phys. Rev. D {\bf 73}, 072003 (2006).
\bibitem{g-2t1} F. Jegerlehner, Acta Phys. Polon. B {\bf 38}, 3021 (2007).
\bibitem{g-2t2} M. Passera, W.J. Marciano and A. Sirlin, arXiv:0809.4062 [hep-ph]  (2008).
\bibitem{ggd0} V.M. Abazov et al. (D0 Collaboration), arXiv:0901.1887 [hep-ex] (2009).
\bibitem{comphep} E.E. Boos et al. (CompHEP Collaboration), Nucl. Instr. Meth. A {\bf 534}, 250 (2004).
\end{thebibliography}
\end{document}